\documentclass[12pt]{article}
\usepackage{graphicx}
\usepackage{amssymb}
\usepackage{cite}

\begin{document}

\def\qq{\langle \bar q q \rangle}
\def\uu{\langle \bar u u \rangle}
\def\dd{\langle \bar d d \rangle}
\def\sp{\langle \bar s s \rangle}
\def\GG{\langle g_s^2 G^2 \rangle}
\def\Tr{\mbox{Tr}}
\def\figt#1#2#3{
        \begin{figure}
        $\left. \right.$
        \vspace*{-2cm}
        \begin{center}
        \includegraphics[width=10cm]{#1}
        \end{center}
        \vspace*{-0.2cm}
        \caption{#3}
        \label{#2}
        \end{figure}
    }

\def\figb#1#2#3{
        \begin{figure}
        $\left. \right.$
        \vspace*{-1cm}
        \begin{center}
        \includegraphics[width=10cm]{#1}
        \end{center}
        \vspace*{-0.2cm}
        \caption{#3}
        \label{#2}
        \end{figure}
                }

\def\ds{\displaystyle}
\def\beq{\begin{equation}}
\def\eeq{\end{equation}}
\def\bea{\begin{eqnarray}}
\def\eea{\end{eqnarray}}
\def\beeq{\begin{eqnarray}}
\def\eeeq{\end{eqnarray}}
\def\ve{\vert}
\def\vel{\left|}
\def\ver{\right|}
\def\nnb{\nonumber}
\def\ga{\left(}
\def\dr{\right)}
\def\aga{\left\{}
\def\adr{\right\}}
\def\lla{\left<}
\def\rra{\right>}
\def\rar{\rightarrow}
\def\nnb{\nonumber}
\def\la{\langle}
\def\ra{\rangle}
\def\ba{\begin{array}}
\def\ea{\end{array}}
\def\tr{\mbox{Tr}}
\def\ssp{{\Sigma^{*+}}}
\def\sso{{\Sigma^{*0}}}
\def\ssm{{\Sigma^{*-}}}
\def\xis0{{\Xi^{*0}}}
\def\xism{{\Xi^{*-}}}
\def\qs{\la \bar s s \ra}
\def\qu{\la \bar u u \ra}
\def\qd{\la \bar d d \ra}
\def\qq{\la \bar q q \ra}
\def\gGgG{\la g^2 G^2 \ra}
\def\q{\gamma_5 \not\!q}
\def\x{\gamma_5 \not\!x}
\def\g5{\gamma_5}
\def\sb{S_Q^{cf}}
\def\sd{S_d^{be}}
\def\su{S_u^{ad}}
\def\sbp{{S}_Q^{'cf}}
\def\sdp{{S}_d^{'be}}
\def\sup{{S}_u^{'ad}}
\def\ssp{{S}_s^{'??}}

\def\sig{\sigma_{\mu \nu} \gamma_5 p^\mu q^\nu}
\def\fo{f_0(\frac{s_0}{M^2})}
\def\ffi{f_1(\frac{s_0}{M^2})}
\def\fii{f_2(\frac{s_0}{M^2})}
\def\O{{\cal O}}
\def\sl{{\Sigma^0 \Lambda}}
\def\es{\!\!\! &=& \!\!\!}
\def\ap{\!\!\! &\approx& \!\!\!}
\def\ar{&+& \!\!\!}
\def\ek{&-& \!\!\!}
\def\kek{\!\!\!&-& \!\!\!}
\def\cp{&\times& \!\!\!}
\def\se{\!\!\! &\simeq& \!\!\!}
\def\eqv{&\equiv& \!\!\!}
\def\kpm{&\pm& \!\!\!}
\def\kmp{&\mp& \!\!\!}
\def\mcdot{\!\cdot\!}
\def\erar{&\rightarrow&}


\def\simlt{\stackrel{<}{{}_\sim}}
\def\simgt{\stackrel{>}{{}_\sim}}



\renewcommand{\textfraction}{0.2}    
\renewcommand{\topfraction}{0.8}

\renewcommand{\bottomfraction}{0.4}
\renewcommand{\floatpagefraction}{0.8}
\newcommand\mysection{\setcounter{equation}{0}\section}

\def\baeq{\begin{appeq}}     \def\eaeq{\end{appeq}}
\def\baeeq{\begin{appeeq}}   \def\eaeeq{\end{appeeq}}
\newenvironment{appeq}{\beq}{\eeq}
\newenvironment{appeeq}{\beeq}{\eeeq}
\def\bAPP#1#2{
 \markright{APPENDIX #1}
 \addcontentsline{toc}{section}{Appendix #1: #2}
 \medskip
 \medskip
 \begin{center}      {\bf\LARGE Appendix #1 }{\quad\Large\bf #2}
\end{center}
 \renewcommand{\thesection}{#1.\arabic{section}}
\setcounter{equation}{0}
        \renewcommand{\thehran}{#1.\arabic{hran}}
\renewenvironment{appeq}
  {  \renewcommand{\theequation}{#1.\arabic{equation}}
     \beq
  }{\eeq}
\renewenvironment{appeeq}
  {  \renewcommand{\theequation}{#1.\arabic{equation}}
     \beeq
  }{\eeeq}
\nopagebreak \noindent}

\def\eAPP{\renewcommand{\thehran}{\thesection.\arabic{hran}}}

\renewcommand{\theequation}{\arabic{equation}}
\newcounter{hran}
\renewcommand{\thehran}{\thesection.\arabic{hran}}

\def\bmini{\setcounter{hran}{\value{equation}}
\refstepcounter{hran}\setcounter{equation}{0}
\renewcommand{\theequation}{\thehran\alph{equation}}\begin{eqnarray}}
\def\bminiG#1{\setcounter{hran}{\value{equation}}
\refstepcounter{hran}\setcounter{equation}{-1}
\renewcommand{\theequation}{\thehran\alph{equation}}
\refstepcounter{equation}\label{#1}\begin{eqnarray}}


\newskip\humongous \humongous=0pt plus 1000pt minus 1000pt
\def\caja{\mathsurround=0pt}


\title{
         {\Large
                 {\bf
 Electric Quadrupole and Magnetic Octupole Moments of the
 Light Decuplet Baryons Within  Light Cone QCD Sum Rules
                 }
         }
      }

\author{\vspace{1cm}\\
{\small T. M. Aliev \thanks {e-mail:
taliev@metu.edu.tr}~\footnote{permanent address:Institute of
Physics,Baku,Azerbaijan}\,\,, K. Azizi \thanks {e-mail:
kazizi@dogus.edu.tr}\,\,, M. Savc{\i} \thanks
{e-mail: savci@metu.edu.tr}} \\
{\small $^{\dag,\S}$Physics Department, Middle East Technical University,
06531 Ankara, Turkey}\\
{\small$^{\ddag}$ Physics Division,  Faculty of Arts and Sciences, Do\u gu\c s University,}\\
{\small Ac{\i}badem-Kad{\i}k\"oy,  34722 Istanbul, Turkey} }

\date{}

\begin{titlepage}
\maketitle
\thispagestyle{empty}

\begin{abstract}
The electric quadrupole  and magnetic octupole  moments of the
light decuplet baryons  are calculated in the framework of the
light cone QCD sum rules. The  obtained  non-vanishing values for the electric quadrupole and magnetic octupole moments of these  baryons show nonspherical charge distribution. The sign of electric quadrupole  moment is positive  for $\Omega^-$, $\Xi^{*-}$, $\Sigma^{*-}$ and negative for $\Sigma^{*+}$, which correspond to the prolate and oblate charge distributions, respectively.  A comparison of the obtained results
with the predictions of  non-covariant quark  model which shows a good consistency between two approaches is also  presented. Comparison of the obtained results on the multipole moments of the decuplet baryons containing strange quark with those of $\Delta$ baryons shows a large SU(3) flavor symmetry breaking.
\end{abstract}

~~~PACS number(s): 11.55.Hx, 13.40.Em, 13.40.Gp
\end{titlepage}


\section{Introduction}

Detailed study of the electromagnetic properties of baryons, such as
electromagnetic multipole moments and electromagnetic form factors,
can give essential information about the nonperturbative structure of QCD.
These multipole moments  are  related to the spatial charge and current
distributions in baryons. Therefore, calculating these parameters could
provide valuable insight on the internal structure as well as the geometric
shape of baryons.
The dominant elastic form factors of decuplet baryons are the charge
$G_{E_0}$ and magnetic dipole $G_{M_1}$. The subdominant form factors are
the electric quadrupole $G_{E_2}$ and magnetic octupole $G_{M_3}$
(all these form factors are defined below). Note that, at $q^2=0$, the form
factors $G_{M_1}$, $G_{E_2}$ and $G_{M_3}$ give the magnetic dipole $\mu_B$, electric
quadrupole ${\cal Q}_B$, and the magnetic octupole moments ${\cal O}_B$,
respectively \cite{R9704}. The size of the higher multipole moments ${\cal Q}_B$ and
${\cal O}_B$ provide information about the
deformation of the baryon and its direction.

 Few words about the experimental prospects for measurement of the multipole moments are in order. There are two types of  transitions for studying the multipole moments of the ground state decuplet baryons: diagonal transitions between them and off diagonal transitions between the decuplet and octet baryons i.e., $\Delta\rightarrow N$, $\Sigma^*\rightarrow \Sigma$, $\Sigma^*\rightarrow \Lambda$ and $\Xi^*\rightarrow \Xi$. The couplings of diagonal decuplet-decuplet-photon transitions,  obviously, can be measured only by virtual photon exchange. The magnetic moment of $\Delta^{+}$ has been measured via $\gamma p\rightarrow \pi^0 \gamma'p$  reaction \cite{ R9702}.  However, measurement of the electric quadrupole by studying the diagonal transition  is practically hopeless. This is due to the fact that the electric quadrupole operator is T-odd quantity and matrix element of this operator between the same initial and final states is equal to zero. Therefore, for the experimental study of the electric quadrupole moment, the suitable place is  off-diagonal transitions. For example, the $E_2$ transition can be measured in reaction $octet ~baryons+X\rightarrow decuplet~ baryons+X$ \cite{Eschrich}, where X is heavy nucleons and also kaon photoproduction experiments $\gamma p\rightarrow K+$ $decuplet$ $\rightarrow K+$ $octet$ $+\gamma$ \cite{CE}. Analysis of the electron-proton and photon-proton scattering
experiments leads to a nonzero quadrupole moment of $p\rightarrow \Delta^{+}$
transition \cite{R9703}.

There are large number of works
in literature which are devoted to the investigation of the magnetic moment
of hadrons, but unfortunately relatively little is known about the other
multipole moments. Therefore, further detailed analysis is needed in studying
higher multipole moments of the hadrons.
Since obtaining direct experimental information about the electromagnetic
multipole moments  of these baryons is very limited, the theoretical  studies play important role in this respect.
The electric quadrupole and magnetic octupole moments of
the $\Delta$ baryons have been calculated within the frame work of the light
cone QCD sum rules (LCSR) in \cite{R9705}. It is obtained that both
quadrupole and magnetic octupole moments have nonzero values and  negative
sign, for example, for $\Delta^{+}$, implying that  the quadrupole and octupole
moment distributions of $\Delta^{+}$   are oblate and have the same geometric
shape as the charge distribution. The same result has also been obtained by
analyzing the quadrupole and octupole moments of $\Delta$  baryons in spectator
quark model \cite{R9706}. These multipole moments for  $\Delta$  baryons have
also been discussed in constituent quark model  with configuration mixing but no
exchange currents (impulse approximation), and constituent quark model  with
exchange currents but no configuration mixing \cite{R9707}.

Present work is devoted to the calculation of the electric quadrupole and
magnetic octupole moments of the decuplet baryons
in the framework of the light
cone QCD sum rules. As has already been noted, these multipole moments of the $\Delta$ baryons have
been calculated in \cite{R9705} in the same framework.   Here, we extend the calculation of the multipole moments
to the other members of the decuplet spin 3/2 baryons, i.e.,
$\Sigma^{*+,0,-}$, $\Xi^{*0,-}$ and $\Omega^{*-}$.
Note that, the magnetic octupole  moments of these baryons have been calculated in
non--covariant quark model (NCQM) \cite{R9708}. Recently, the electromagnetic
form factors of decuplet baryons have been calculated in lattice QCD in
\cite{R9709}. Here, also we stress that the magnetic moments of the decuplet baryons have been studied in \cite{savci} whithin light cone QCD sum rules. The main difference between the present study and \cite{savci} is that, in the present work we calculate additional form factors corresponding to different kinematical structures which are related to the higher multipole moments such as quadrupole and octupole.
The outline of the paper is as follows:
in section II, the light cone QCD sum rules for the electromagnetic form factors
are obtained in LCSR. Section III encompasses the numerical analysis of the
form factors, a comparison of the results with the predictions of the other
approaches and discussion.

\section{Light cone QCD sum rules for electric quadrupole and magnetic octupole
moments  of the Decuplet baryons  }
For study the properties of hadrons in the  sum rule formalism, the  main working tool is the correlation function. To calculate   the multipole form factors of
the decuplet baryons, we consider the following correlation function:
\begin{equation}\label{T}
T_{\mu\nu}=i\int d^{4}xe^{ipx}\langle0\mid T\{\eta_{\mu}(x)\bar{\eta}_{\nu}(0) \}
\mid0\rangle_{\gamma}~,
\end{equation}
where $\eta_\mu$ is the interpolating current for the decuplet baryons,
and  $\gamma$ denotes  the electromagnetic field. In the sum rule method, the
above--mentioned correlation function is calculated in two different ways:
on the phenomenological or physical side, it is saturated by a tower of baryons with
the same  quantum numbers as their interpolating current. On the  QCD or theoretical
side, it  is calculated using the operator product expansion (OPE), where the
short-- and long--distance quark--gluon interactions are separated. The former
is calculated using QCD perturbation theory, whereas the latter are parameterized
in terms of the light-cone distribution amplitudes of the photon in light cone version of QCD sum rules.
The  electromagnetic form factors  are determined by matching  these two
representations of the correlation function.

First, let us calculate the  physical part of the correlation function.
By isolating the contributions of the ground state baryons from  Eq. (\ref{T}), we obtain
\begin{eqnarray}
\label{T2}
T_{\mu\nu}&=&\frac{\langle0\mid \eta_{\mu}\mid
B(p_{2})\rangle}{p_{2}^{2}-m_{B}^{2}}\langle B(p_{2})\mid
B(p_{1})\rangle_\gamma\frac{\langle B(p_{1})\mid
\bar{\eta}_{\nu}\mid 0\rangle}{p_{1}^{2}-m_{B}^{2}} + \cdots~,
\end{eqnarray}
where $p_{1}=p+q$,  $p_{2}=p$ and q is the momentum of a photon.
The  dots mean contributions of the higher states and continuum.

It follows from Eq. (\ref{T2}) that, for calculation of the physical
part, we need to know the matrix element of the interpolating current between
the vacuum and the decuplet baryon state as well as transition matrix element, $\langle
B(p_{2})\mid B(p_{1})\rangle_\gamma$. The  $\langle0\mid \eta_{\mu}(0)\mid
B(p,s)\rangle$ is defined in terms of the residue of the corresponding
decuplet baryons,
$\lambda_{B}$ as:
\begin{equation}
\label{lambdabey}
\langle0\mid \eta_{\mu}(0)\mid B(p,s)\rangle=\lambda_{B}u_{\mu}(p,s),
\end{equation}
where  $u_{\mu}(p,s)$ is the
Rarita-Schwinger spinor. The transition  matrix element $\langle
B(p_{2})\mid B(p_{1})\rangle_\gamma$ can be parameterized in terms of four
form factors as
\cite{R9706,R9710,R9711}:
\begin{eqnarray}\label{matelpar}
\langle B(p_{2})\mid B(p_{1})\rangle_\gamma \es -e\bar
u_{\mu}(p_{2})\left\{\vphantom{\int_0^{x_2}}F_{1}g^{\mu\nu}\not\!\varepsilon-
\frac{1}{2m_{B}}\left
[F_{2}g^{\mu\nu}+F_{4}\frac{q^{\mu}q^{\nu}}{(2m_{B})^2}\right]
\not\!\varepsilon\not\!q \right. \nnb \\
\ar \left. F_{3}\frac{1}{(2m_{B})^2}q^{\mu}q^{\nu}\not\!\varepsilon
\vphantom{\int_0^{x_2}}\right\}
u_{\nu}(p_{1}),\nonumber\\
\end{eqnarray}
where $\varepsilon$ is the polarization vector of the photon  and  $F_{i}$
are form factors as functions of transfer momentum square $q^2=(p_{1}-p_{2})^2$. For obtaining  the
expression for  the correlation function from physical side,  summation over spins of the spin
3/2 particles is performed using
\begin{equation}\label{raritabela}
\sum_{s}u_{\mu}(p,s)\bar u_{\nu}(p,s)=(\not\!p+m_{B})
\left\{-g_{\mu\nu}
+\frac{1}{3}\gamma_{\mu}\gamma_{\nu}-\frac{2p_{\mu}p_{\nu}}
{3m^{2}_{B}}-\frac{p_{\mu}\gamma_{\nu}-p_{\nu}\gamma_{\mu}}{3m_{B}}\right\}.
\end{equation}
 In principle, using the above equations, we can obtain the final
expression of the physical side of the correlation function, but we come across
with two difficulties.
a) Not only spin 3/2, but spin
1/2 particles   contribute to the correlation function, i.e., the matrix element
of the   current $\eta_{\mu}$ of the spin 3/2 particles between vacuum and spin 1/2
states is nonzero. This matrix element in general form can be written as
\begin{equation}\label{spin12}
\langle0\mid \eta_{\mu}(0)\mid B(p,s=1/2)\rangle=(A  p_{\mu}+B\gamma_{\mu})u(p,s=1/2).
\end{equation}
Using the condition $\gamma_\mu \eta^\mu = 0$, one can
immediately obtain that $B=-\frac{A}{4}m$. b) All Lorentz structures are not
independent (for more details, see \cite{R9712}).

In order to eliminate  the unwanted spin 1/2 contributions and obtain only
independent structures, the  ordering procedure of Dirac matrices are applied and in the present work, we choose it as
$\gamma_{\mu}\not\!p\not\!\varepsilon\not\!q\gamma_{\nu}$. After this ordering procedure,
we obtain the final expression of the physical side of the correlation function as follows:
\begin{eqnarray}
\label{final phenpart}
T_{\mu\nu} &=& \lambda_{_{B}}^{2}{1\over (p_{1}^{2}-m_{_{B}}^{2})(p_{2}^{2}-m_{_{B}}^{2})}
\left[\vphantom{\int_0^{x_2}} 2 (\varepsilon\cdot p)
g_{\mu\nu} \not\!p F_1 + {1\over m_B} (\varepsilon\cdot p) g_{\mu\nu} \not\!p
\not\!q F_2 \right. \nnb \\
\ar \left. {1\over 2 m_B^2} (\varepsilon\cdot p)
q_\mu q_\nu \not\!p F_3
+ {1\over 4 m_B^2}(\varepsilon \cdot p) q_\mu q_\nu \not\!q F_4
+ \mbox{other independent structures } \right. \nnb \\
\ar \left. \mbox{ structures with }\gamma_{\mu}~
\mbox{at the beginning and }\gamma_{\nu}\mbox{ at the end or which are } \right. \nnb \\
&&\left. \mbox{ proportional to }
p_{2\mu}\mbox{ or }p_{1\nu}
\vphantom{\int_0^{x_2}}\right].
\end{eqnarray}
For calculation of   the four form factors,  we need four structures. We will choose the structures
$(\varepsilon.p)g_{\mu\nu}\not\!p$,  $(\varepsilon.p)g_{\mu\nu}\not\!p\not\!q$,
$(\varepsilon.p)q_{\mu}q_{\nu}\not\!p$ and $(\varepsilon.p)q_{\mu}q_{\nu}\not\!q$
for determination of the form factors $F_1$, $F_2$, $F_3$ and $F_4$, respectively. In the experiments the multipole form factors, $G_{E_0}$ (charge), $G_{M_1}$ (magnetic dipole), $G_{E_2}$ (electric quadrupole) and $G_{M_3}$ (magnetic octupole)  are usually measured. Therefore, we need relations between two sets of form factors.
The multipole form factors are
defined in terms of the form factors $F_{i}(q^2)$ as \cite{R9706,R9710,R9711,R9713}:
\begin{eqnarray}
G_{E_0}(q^2) &=& \left[ F_1(q^2) -x F_2(q^2) \right]
\left( 1+ {2\over 3}x \right)  - \left[ F_3(q^2) -x F_4(q^2) \right]
{x \over 3}\left( 1 + x \right)~, \nonumber \\
G_{M_1}(q^2) &=& \left[ F_1(q^2) +F_2(q^2) \right]
\left( 1+ {4\over 5}x \right) - {2\over 5} \left[ F_3(q^2)  +
F_4(q^2)\right] x\left( 1 + x \right)~, \nonumber \\
G_{E_2}(q^2) &=& \left[ F_1(q^2) -x F_2(q^2) \right]  -
\frac{1}{2}\left[ F_3(q^2) -x F_4(q^2)
\right] \left( 1+ x \right)~,  \nonumber \\
 G_{M_3}(q^2) &=&
\left[ F_1(q^2) + F_2(q^2)\right] -\frac{1}{2} \left[ F_3(q^2)  +
F_4(q^2)\right] \left( 1 + x \right)~,
\end{eqnarray}
where $x = -q^2/4m_B^2$.
At $q^2=0$, we obtain
\begin{eqnarray}\label{mqo1}
G_{M_1}(0) &=& F_1(0) +F_2(0) ~,\nonumber\\
G_{E_2}(0)&=&F_{1}(0)-\frac{1}{2}F_{3}(0)~,\nonumber\\
G_{M_3}(0)&=&F_{1}(0)+F_{2}(0)-\frac{1}{2}[F_{3}(0)+F_{4}(0)]~.
\end{eqnarray}
The magnetic dipole $\mu_B$, the  electric
quadrupole ${\cal Q}_{B}$, and the magnetic octupole ${\cal O}_{B}$ moments
are defined in terms of these form factors at $q^2=0$ in the following way:
 \begin{eqnarray}\label{mqo2}
\mu_B &=& {e\over 2 m_B} G_{M_1}(0)~, \nonumber\\
{\cal Q}_{B}&=&\frac{e}{m_{B}^2}G_{E_2}(0)~, \nonumber\\
{\cal O}_{B}&=&\frac{e}{2m_{B}^3}G_{M_3}(0)~.
\end{eqnarray}

The QCD side  of the correlation function, on the other hand, can be calculated
by the help of  the OPE in deep Euclidean region where
$p^2\ll0$ and $(p+q)^2\ll0$.
For this aim we need to know  the explicit expressions of the interpolating
currents of the corresponding baryons.  The interpolating currents for decuplet baryons are \cite{decupc}
\begin{eqnarray}\label{currentguy}
\eta_\mu^{\Sigma^{*0}} &=&\sqrt{\frac{2}{3}} \epsilon^{abc}
[ (u^{aT} C \gamma_\mu d^b) s^c +
(d^{aT} C  \gamma_\mu s^b) u^c + (s^{aT} C \gamma_\mu u^b) d^c]~,\nonumber\\
\eta_\mu^{\Sigma^{*+}} &=&\frac{1}{\sqrt{2}}\eta_\mu^{\Sigma^{*0}}
(d\rightarrow u)~,\nonumber\\
\eta_\mu^{\Sigma^{*-}} &=&\frac{1}{\sqrt{2}}\eta_\mu^{\Sigma^{*0}}
(u\rightarrow d)~,\nonumber\\
\eta_\mu^{\Xi^{*0}} &=&\eta_\mu^{\Sigma^{*-}} (s\rightarrow u)
(d\rightarrow s)~,\nonumber\\
\eta_\mu^{\Xi^{*-}} &=&\eta_\mu^{\Xi^{*0}} (u\rightarrow d)~,\nonumber\\
\eta_\mu^{\Omega^{*-}} &=&\frac{1}{\sqrt{3}}\eta_\mu^{\Sigma^{*+}} (u\rightarrow
s)~,\nonumber\\
\end{eqnarray}
where   $a$, $b$ and $c$ are color
indices and C is the charge conjugation operator. After contracting out all
quark pairs in  Eq. (\ref{T}) using the
Wick's theorem, we obtain the following expression for the correlation
function of the $\Sigma^{*0}\rightarrow\Sigma^{*0}\gamma$ transition  in terms
of the  quark propagators:
\begin{eqnarray}
\label{tree expresion.m}
\Pi^{\Sigma^{*0}\rightarrow\Sigma^{*0}\gamma}_{\mu \nu} &=& -
\frac{2i}{3}\epsilon_{abc}\epsilon_{a'b'c'}\int
d^{4}xe^{ipx}\langle\gamma(q)\mid\{S_{d}^{ca'}
\gamma_{\nu}S'^{bb'}_{u}\gamma_{\mu}S_{d}^{ac'}\nonumber\\&+&S_{d}^{cb'}
\gamma_{\nu}S'^{aa'}_{s}\gamma_{\mu}S_{u}^{bc'}+ S_{s}^{ca'}
\gamma_{\nu}S'^{bb'}_{d}\gamma_{\mu}S_{u}^{ac'}+S_{s}^{cb'}
\gamma_{\nu}S'^{aa'}_{u}\gamma_{\mu}S_{d}^{bc'}\nonumber\\&+& S_{u}^{cb'}
\gamma_{\nu}S'^{aa'}_{d}\gamma_{\mu}S_{s}^{bc'}
+ S_{u}^{ca'}\gamma_{\nu}S'^{bb'}_{s}\gamma_{\mu}S_{d}^{ac'}
\nonumber\\&+& Tr(\gamma_{\mu}S_{s}^{ab'}\gamma_{\nu}S'^{ba'}_{u})S^{cc'}_{d}
+Tr(\gamma_{\mu}S_{u}^{ab'}\gamma_{\nu}S'^{ba'}_{d})S^{cc'}_{s}
\nonumber\\&+& Tr(\gamma_{\mu}S_{d}^{ab'}\gamma_{\nu}S'^{ba'}_{s})S^{cc'}_{u}\}
\mid 0\rangle,
\end{eqnarray}
where $S'=CS^TC$ and $S_{u,d,s}$ are the light quark
propagators. The correlation
functions for other transitions can be obtained by the replacements mentioned in
Eq. (\ref{currentguy}). The expression of the light quark propagator in the external field
is calculated in \cite{R9714,R9715}:
\begin{eqnarray}\label{onuc}
\label{Slight}
S_q(x) &=& S^{free} (x) - {\langle qq\rangle \over 12} \left(1 -i {m_q \over
4} \not\!{x} \right) - {x^2 \over 192} m_0^2 \langle qq\rangle
\left(1 -i {m_q \over 6} \not\!{x} \right) \nnb \\
\ek i g_s \int_0^1 du \left\{
{\not\!x \over 16 \pi^2 x^2} G_{\mu\nu}(ux) \sigma^{\mu\nu}
- u x^\mu G_{\mu\nu}(ux) \gamma^\nu {i \over 4 \pi^2 x^2} \right. \nnb \\
\ek \left. i {m_q \over 32 \pi^2} G_{\mu\nu}(ux) \sigma^{\mu\nu} \left[ \ln\left( -{x^2
\Lambda^2 \over 4} + 2 \gamma_E \right) \right] \right\}~,
\end{eqnarray}
where $\Lambda$ is the scale parameter, and following \cite{R9716}, we choose
it at the factorization scale $\Lambda=0.5~GeV-1.0~GeV$. The correlation function contain three pieces:
a) Short distance contributions,
b) ``Mixed" contributions,
c) Large distance contributions when a photon is radiated at long distance.

Different terms in Eq. (\ref{onuc}) give contributions to the different pieces of the correlation function.
The short distance contributions can easily
be obtained from Eq. (\ref{tree expresion.m}) by replacing one of the
propagators by,
\begin{eqnarray}\label{ondort}
\label{shortDC}
S_{\alpha\beta}^{ab} = \left\{ \int d^4y S^{free} (x-y) \not\!\!{A}  S^{free} (y)
\right\}_{\alpha\beta}^{ab}~,
\end{eqnarray}
where $S^{free}$ is the light quark propagator given as,
\begin{eqnarray}\label{onbes}
\label{Sfree}
S^{free} (x) = {i \not\!x \over 2 \pi^2 x^4} - {m_q \over 4 \pi^2 x^2}~.
\end{eqnarray}
 and two other quark propagators are replaced by the free quark propagator.

In the ``mixed" contributions case, a photon interacts with quark fields perturbatively. Therefore, one of the quark propagators is replaced by Eq. (\ref{ondort}), two other propagators either  both are  replaced by
\begin{eqnarray}\label{onalti}
S_{\alpha\beta}^{ab} = - {1\over 4} \bar{q}^a \Gamma_j q^b
(\Gamma_j)_{\alpha\beta}~,
\end{eqnarray}
which both can form quark condensates, or one of them is replaced by Eq. (\ref{onalti}) and second one by the free quark propagator. In  Eq. (\ref{onalti}),  $\Gamma_j$ is the full set of Dirac matrices.

The large distance contributions can be obtained from  Eq. (\ref{tree expresion.m}) by following replacements: One of the quark propagators is replaced by Eq. (\ref{onalti}) and a photon interacts with the quark fields at large distance, i.e.,  the matrix elements  of the nonlocal
operators $\bar{q}(x_1) \Gamma q^\prime (x_2)$ and $\bar{q}(x_1)
G_{\mu\nu}\Gamma q^\prime (x_2)$ appear between the vacuum and the vector meson
states, which is parameterized in terms of photon distribution amplitudes (DA's). Two other propagators are both replaced by free quark propagator, or one of them is replaced by free quark propagator and second one is replaced  by Eq. (\ref{onalti}) and then it interact with QCD vacuum, i.e., it forms a quark condensate, or both of propagators are replaced by Eq. (\ref{onalti}) and then they form quark condensates.

Using the expressions of the  light propagators and the photon
DA's and separating the coefficient of the structures mentioned before and
applying double Borel transformation with respect to the variables $p_{2}^2=p^2$ and
$p_{1}^2=(p+q)^2$  to suppress the contributions of the higher states and continuum,
sum rules for the form factors  $F_{1}$, $F_{2}$,  $F_{3}$ and $F_{4}$ are obtained.
The explicit expressions of the sum rules for these form factors are given in the   Appendix--A.
From the expressions of the form factors it is clear that, to obtain form factors,
we need to know the explicit expressions of residues of the corresponding baryons.
The explicit expressions for these residues  are given in \cite{R9717,R9718}.

\section{Numerical analysis}

Present  section is devoted to the numerical analysis for the,
electric quadrupole and magnetic octupole  moments
of the light spin 3/2 baryons. The values for input parameters used in
the analysis of the sum rules for the $F_{1}$, $F_{2}$, $F_{3}$ and
$F_{4}$ are : $\langle \bar uu(1~GeV)\rangle  = \langle \bar dd(1~GeV) \rangle = -(1.65\pm0.15)\times10^{-2}~GeV^3$ \cite{BL},
$\langle \bar ss(1~GeV) \rangle  = 0.8 \langle \bar uu(1~GeV)\rangle$,  $ m_{s}(2~GeV)=(111 \pm 6)~MeV$ at  $\Lambda_{QCD}=330~MeV$ \cite{Dominguez}, $m_0^2(1~GeV) = (0.8\pm0.2)~GeV^2$
\cite{R9718} and $f_{3 \gamma} = - 0.0039~GeV^2$ \cite{R9719}. The
value of the magnetic susceptibility is taken to be
$\chi(1~GeV)=-3.15\pm0.3~GeV^{-2}$  \cite{R9719}. As has already be noted,
the main input parameters in light cone sum rules are the DA's. The explicit
expression of the photon DA's are given in \cite{R9719}.

The sum rules for the electromagnetic form factors  also contain two auxiliary
parameters: Borel mass parameter $M^2$ and continuum threshold $s_{0}$. The
physical quantities  should be independent of these
parameters. Therefore, we look for a region for these parameters such that the
electromagnetic form factors are independent of them.   The working region for
$M^2$ are found requiring that not only  the contributions of the
higher states and continuum should be less than the ground state
contribution, but the highest power of $1/M^{2}$ be less than  say
$30^0/_{0}$ of the highest power of $M^{2}$. These conditions are satisfied in the regions
$1.1~GeV^2\leq M^{2}\leq1.6~GeV^2 $, $1.2~GeV^2\leq M^{2}\leq1.7~GeV^2 $ and $1.4~GeV^2\leq M^{2}\leq2.4~GeV^2 $ for $\Sigma^{*}$, $\Xi^{*}$ and $\Omega^{*}$ baryons, respectively. In the numerical analysis,
$s_0=(m_B+0.5)^2~GeV^2$ has been used for   value of the continuum threshold.

Our final  results on the  electric quadrupole $Q_{B}$ and magnetic octupole
${\cal O}_{B}$ moments of decuplet baryons are presented in Table 1. The quoted errors in Table 1 can be attributed to the uncertainties
in the variation of the Borel parameter $M^2$, the continuum threshold
$s_0$, as well as the uncertainties in the determination of the other input
parameters entering  the sum rules.
A comparison of our predictions on magnetic octupole moment with the results
obtained in NCQM is also presented in Table 1.
\begin{table}[h]
\renewcommand{\arraystretch}{1.5}
\addtolength{\arraycolsep}{3pt}

$$
\begin{array}{|c|r@{\pm}l|r@{\pm}l|r@{\div}l|}                                  \hline \hline
 & \multicolumn{2}{c|}{\mbox{Quadrupole}~{\cal Q}(fm^2)}
 & \multicolumn{4}{c|}{\mbox{Octupole}~{\cal O}(fm^3)  }                    \\
 & \multicolumn{2}{c|}{\mbox{Present Work}}
 & \multicolumn{2}{c} {\mbox{Present Work}}
 & \multicolumn{2}{c|}{\mbox{NCQM \cite{R9708}}}                            \\ \hline
   \Omega^-          &~~~~~~~ 0.12&0.04    &   0.016&0.004  &  0.003& 0.012 \\
   \Sigma^{\ast -}   &        0.03&0.01    &   0.013&0.004  &  0.008&0.012 \\
   \Sigma^{\ast 0}   &      0.0012&0.0004  & -0.001&0.0003 &  0.000&0.002 \\
   \Sigma^{\ast +}   &      -0.028&0.009   &  -0.015&0.005  & -0.004&- 0.012\\
   \Xi^{\ast -}      &        0.045&0.015    &   0.020&0.006  &0.005&0.012  \\
   \Xi^{\ast 0}      &      0.0025&0.0008  & -0.0014&0.0005 &0.000& 0.002     \\ \hline \hline
\end{array}
$$

\caption{Results of the electric quadrupole moment (in units of $fm^2$) and
magnetic octupole moments (in units of $fm^3$) of the decuplet baryons.}
\renewcommand{\arraystretch}{1}
\addtolength{\arraycolsep}{-1.0pt}
\end{table}
 The results for magnetic octupole moments show a  good consistency between our predictions
and
those of the NCQM \cite{R9708}. As has already been noted, the
electromagnetic form factors of the decuplet baryons have been calculated at
$q^2\neq 0$ in \cite{R9709}, so we can not compare our results with theirs.
However, the order of magnitude of our results are in good agreement with
their predictions at low $q^2$.  Comparison between our results on electric quadrupole and magnetic octupole moments of
   $\Sigma^{*+,-}$, $\Xi^{*-}$, $\Omega^-$ and the predictions of  \cite{R9705} for  $\Delta$ baryons, shows a large SU(3) flavor symmetry braking. This violation is larger for $\Xi^{*-}$ and  $\Omega^-$ baryons which contain two and three strange quarks, respectively. In the case of the strange baryons, the results are very sensitive  to the strange quark mass. This sensitivity together with  the different working regions of Borel mass parameter, $M^2$ , and continuum threshold, $s_0$ lead to the different values of multipole moments for $\Sigma^{*}$, $\Xi^{*}$, $\Omega$ baryons.

In conclusion, the  electric quadrupole  and magnetic octupole  moments of
decuplet baryons were calculated  in the framework of the light cone QCD sum
rules. We obtained  non-vanishing values for the electric quadrupole and magnetic octupole moments of these  baryons which mean nonspherical charge distribution. The sign of electric quadrupole  moment is positive  for $\Omega^-$, $\Xi^{*-}$, $\Sigma^{*-}$ and negative for $\Sigma^{*+}$, which correspond to the prolate and oblate charge distributions, respectively. The obtained results are in good consistency with the predictions of  the non-covariant quark model. Comparison of the obtained results on the multipole moments of the decuplet baryons containing strange quark with those of $\Delta$ baryons presents a large SU(3) flavor symmetry breaking.

\section{Acknowledgment}
We thank A. Ozpineci for his useful discussions.

\newpage

\bAPP{A}{}

In this appendix, we present the    sum rules for the
form factors, $F_{1}(0)$, $F_{2}(0)$, $F_{3}(0)$ and $F_{4}(0)$.

\baeeq
\label{eocap01}
F_{1}(q^2=0) \es {1 \over 2 \lambda_{\Sigma^{*0}}^{2}}
e^{m_{\Sigma^{*0}}^{2}/M^{2}}\Bigg\{
{1\over 40 \pi^4} (e_u+e_d+e_s) M^6 \nnb \\
\ek {1\over 6 \pi^2} M^2 m_s \Big[ 2 (e_d+e_s) \uu + 2 (e_u+e_s) \dd -
(e_u+e_d) \sp
\Big] \nnb \\
\ek {1\over 36 \pi^2 M^2} m_s \GG (e_u \dd + e_d \uu)
\Bigg( \gamma_E + \ln {\Lambda^2 \over M^2} \Bigg)  \nnb \\
\ar {1\over 54 \pi^2 M^2} m_s \GG (e_u \dd + e_d \uu) -
{4\over 9 M^2} m_0^2 (e_u \dd \sp + e_d \uu \sp + e_s \uu \dd) \nnb \\
\ek {1\over 144 \pi^2 M^4} m_0^2 m_s \GG (e_u \dd + e_d \uu) \nnb \\
\ar {1\over 54 \pi^2} m_0^2 m_s \Big[ (9 e_d + 10 e_s) \uu +
(9 e_u + 10 e_s) \dd - 4 (e_u + e_d) \sp \Big] \nnb \\
\ar {8\over 9} (e_u \dd \sp + e_d \uu \sp + e_s \uu \dd) \Bigg\}~, \\ \nnb
\\
F_{2}(q^2=0) \es {m_{\Sigma^{*0}} \over \lambda_{\Sigma^{*0}}^{2}}e^{m_{\Sigma^{*0}}^{2}/M^{2}}
\Bigg\{ - {1\over 1152 \pi^4 M^2} m_s \Bigg( \gamma_E + \ln {\Lambda^2 \over M^2}
\Bigg) \nnb \\
\cp \Big\{ [3 M^2 + 2 \pi^2 f_{3\gamma}
\psi_a(u_0) ] \GG (e_u + e_d) + 24 e_s M^6 \Big\} \nnb \\
\ek {1\over 288 \pi^4} M^4 [ m_s (3 e_u + 3 e_d + 11 e_s) +
8 \pi^2 \chi (e_u \uu + e_d \dd + e_s \sp)\phi_\gamma(u_0) ] \nnb \\
\ar {1 \over 144 \pi^2} M^2 \Big\{
6 [ (e_u+e_d) \sp + (e_u+e_s) \dd +(e_d+e_s) \uu] \nnb \\
\ar (e_u \uu + e_d \dd + e_s \sp) \Big(
3 \Bbb{A}(u_0) - 4 [i_2({\cal S},1) + i_2(\widetilde{\cal S},3-4 v) \nnb \\
\ar i_2({\cal T}_2,1-2 v) + 2 i_2({\cal T}_3,1-2 v) -
i_2({\cal T}_4,1-2 v) + 8 \widetilde{\widetilde{i}}_3(h_\gamma) ] \Big) \nnb \\
\ek 3 m_s f_{3\gamma} (e_u+e_d) \psi_a(u_0) \Big\} \nnb \\
\ek {1 \over 54 M^2} m_s \sp (e_u \uu + e_d \dd) [
i_2({\cal S},1) + i_2(\widetilde{\cal S},3-4 v) +
i_2({\cal T}_2,1-2 v) \nnb \\
\ar 2 i_2({\cal T}_3,1-2 v) -
i_2({\cal T}_4,1-2 v) ] +
{1 \over 36 M^2} m_s \Big[ 2 (e_u \dd + e_d \uu) \sp \nnb \\
\ar (e_u \uu + e_d \dd) \sp \Bbb{A}(u_0) +
16 (e_u + e_d) \uu \dd \widetilde{\widetilde{i}}_3(h_\gamma) \Big] \nnb \\
\ar {1\over 864 \pi^2 M^2} m_s f_{3\gamma} \GG (e_u+e_d) \psi_a(u_0) +
{1\over 648 M^2} m_0^2 \Big\{ 24 m_s \sp \chi (e_u \uu + e_d \dd)
\phi_\gamma(u_0) \nnb \\
\ek 11 f_{3\gamma} [(e_d+e_s) \uu + (e_u+e_s) \dd +
(e_u+e_d) \sp] \psi_a(u_0) \Big\} \nnb \\
\ar {2 \over 81 M^4} m_0^2 m_s \Big[(e_u + e_d) \uu \dd - 2 (e_u \uu +
e_d \dd) \sp \Big] \widetilde{\widetilde{i}}_3(h_\gamma) \nnb \\
\ar {1\over 54 M^6} m_s \GG (e_u+e_d) \uu \dd
\widetilde{\widetilde{i}}_3(h_\gamma) \nnb \\
\ar {1\over 108 M^8} m_s m_0^2 \GG (e_u+e_d) \uu \dd
\widetilde{\widetilde{i}}_3(h_\gamma) \nnb \\
\ek {1\over 1152 \pi^4} m_s \GG (e_u + e_d) -
{5\over 432 \pi^2} m_0^2 [ (e_d+e_s) \uu + (e_u+e_s) \dd \nnb \\
\ar (e_u+e_d) \sp ] +
{1\over 18} \Big\{ - 2 m_s \chi (e_u \uu + e_d \dd) \sp \phi_\gamma(u_0) +
f_{3\gamma} [ (e_d+e_s) \uu \nnb \\
\ar (e_u+e_s) \dd + (e_u+e_d) \sp ] \psi_a(u_0)
\Big\} \\ \nnb \\
 F_{3}(q^2=0)&=&{2 m^{2}_{\Sigma^{*0}} \over \lambda^{2}_{\Sigma^{*0}}}e^{m^{2}_{\Sigma^{*0}}/M^{2}}
\Bigg\{
- {7 \over 960 \pi^4} M^4 (e_u+e_d+e_s) \nnb \\
\ek {1 \over 36 \pi^2} M^2 f_{3\gamma} (e_u+e_d+e_s) \Big(
[2 i_2({\cal A},5-4v) + 4 i_2({\cal V},1-2v) -\psi_a(u_0) ]
\Big) \nnb \\
\ar {1\over 36 \pi^2 M^4} m_s (e_u \uu + e_d \dd)
\Big(4 M^4 [5 i_1({\cal T}_1+{\cal T}_2,1) - 3 i_1({\cal T}_3
+ {\cal T}_4,1)] \nnb \\
\ek \GG \widetilde{\widetilde{i}}_3(h_\gamma) \Big)
\Bigg( \gamma_E + \ln {\Lambda^2 \over M^2} \Bigg) \nnb \\
\ar {1 \over 54 M^2 \pi^2} m_0^2 m_s \sp (e_u+e_d)
+ {4 \over 27 M^2} [(e_u+e_d) \uu \dd + (e_u+e_s) \uu \sp \nnb \\
\ar (e_d+e_s) \dd \sp][i_1(3 {\cal T}_1 + 4 {\cal T}_2 - {\cal T}_4,1) +
6 \widetilde{\widetilde{i}}_3(h_\gamma)] \nnb \\
\ek {1 \over 27 M^2} m_s f_{3\gamma} \sp (e_u+e_d)
[i_2({\cal A},5-4v) + 2 i_2({\cal V},1-2v) - 3 \psi_a(u_0) ] \nnb \\
\ar {1 \over 216 M^4 \pi^2} m_s \GG (e_u \uu + e_d \dd) [
i_1(3 {\cal T}_1 + 4 {\cal T}_2 - {\cal T}_4,1) +
4 \widetilde{\widetilde{i}}_3(h_\gamma)] \nnb \\
\ek {2 \over 81 M^4} m_0^2 \Big( 10 [(e_u+e_d) \uu \dd + (e_u+e_s) \uu \sp
+(e_d+e_s) \dd \sp] \widetilde{\widetilde{i}}_3(h_\gamma) \nnb \\
\ar m_s f_{3\gamma} (e_u+e_d) \sp \psi_a(u_0) \Big) \nnb \\
\ek {1\over 72 \pi^2} m_s \Big( 3 (e_u + e_d) \sp -
8 (e_u \uu + e_d \dd) [i_1(2 {\cal T}_1 + {\cal T}_2 - 3 {\cal T}_3 -
2 {\cal T}_4,1) \nnb \\
\ek 3 \widetilde{\widetilde{i}}_3(h_\gamma)] \Big) \\ \nnb \\
F_{4}(q^2=0) \es {4m^{2}_{\Sigma^{*0}} \over \lambda^{2}_{\Sigma^{*0}}}e^{m^{2}_{\Sigma^{*0}}/M^{2}}
\Bigg\{ - {1 \over 160 \pi^4} M^4 (e_u+e_d+e_s) \nnb \\
\ek {1 \over 144 \pi^2} M^2 f_{3\gamma} (e_u+e_d+e_s) \Big(
4 [4 i_2({\cal A},1+v) + 4 i_2({\cal V},1-v) +
\widetilde{i}_3(\psi_v) ] - 3 \psi_a(u_0) \Big) \nnb \\
\ar {1\over 72 M^4 \pi^2} m_s (e_u \uu + e_d \dd)
\Big[16 M^4 i_1({\cal T}_1+{\cal T}_2-{\cal T}_3
 - {\cal T}_4,1) - \GG \widetilde{\widetilde{i}}_3(h_\gamma) \Big] \nnb \\
\cp \Bigg( \gamma_E + \ln {\Lambda^2 \over M^2} \Bigg) + {1 \over 108 M^2
\pi^2} m_0^2 m_s \sp (e_u+e_d) \nnb \\
\ar {4 \over 27 M^2} [(e_u+e_d) \uu \dd + (e_u+e_s) \uu \sp \nnb \\
\ar (e_d+e_s) \dd \sp][4 i_1({\cal T}_2 - {\cal T}_4 ,v) +
3 \widetilde{\widetilde{i}}_3(h_\gamma)] \nnb \\
\ek {1 \over 108 M^2} m_s f_{3\gamma} \sp (e_u+e_d)
[8 i_2({\cal A},1+v) + 8 i_2({\cal V},1-v) +
12 \widetilde{i}_3(\psi_v) - 9 \psi_a(u_0) ] \nnb \\
\ar {1 \over 108 M^4 \pi^2} m_s \GG (e_u \uu + e_d \dd) [
2 i_1({\cal T}_2 - {\cal T}_4,v) +
\widetilde{\widetilde{i}}_3(h_\gamma)] \nnb \\
\ek {1 \over 81 M^4} m_0^2 \Big( 10 [(e_u+e_d) \uu \dd + (e_u+e_s) \uu \sp
+(e_d+e_s) \dd \sp] \widetilde{\widetilde{i}}_3(h_\gamma) \nnb \\
\ar m_s f_{3\gamma} (e_u+e_d) \sp \psi_a(u_0) \Big) \nnb \\
\ek {1\over 72 \pi^2} m_s \Big( 3 (e_u + e_d) \sp -
4 (e_u \uu + e_d \dd) [4 i_1({\cal T}_1 - {\cal T}_3,1) \nnb \\
\ar 4 i_1({\cal T}_2-{\cal T}_4,1-2 v) -
3 \widetilde{\widetilde{i}}_3(h_\gamma)] \Big) \Bigg\}
\eaeeq
where,
the Borel parameter $M^2$  is defined as $M^{2}=M_{1}^{2}M_{2}^{2}/M_{1}^{2}+M_{2}^{2}$ and
$u_{0}=M_{1}^{2}/(M_{1}^{2}+M_{2}^{2})$.  Since the masses of
the initial and final baryons are the same, we have set $M_{1}^{2}=M_{2}^{2}$ and $u_{0}=1/2$.
The continuum subtractions have been made via $M^{2n}\rightarrow M^{2n}E_n(x) $,
where $E_n(x)=1-e^{-x}\sum_{i=0}^{n-1}\frac{x^i}{i!}$ with $x=s_0/M^2$.

The functions $i_n$, $\widetilde{i}_3$ and
$\widetilde{\widetilde{i}}_3$ are also defined as
\baeeq
\label{nolabel}
i_1(\phi,f(v)) \es \int {\cal D}\alpha_i \int_0^1 dv
\phi(\alpha_{\bar{q}},\alpha_q,\alpha_g) f(v) \theta(k-u_0)~, \nnb \\
i_2(\phi,f(v)) \es \int {\cal D}\alpha_i \int_0^1 dv
\phi(\alpha_{\bar{q}},\alpha_q,\alpha_g) f(v) \delta(k-u_0)~, \nnb \\
\widetilde{i}_3(f(u)) \es \int_{u_0}^1 du f(u)~, \nnb \\
\widetilde{\widetilde{i}}_3(f(u)) \es \int_{u_0}^1 du (u-u_0) f(u)~,
\eaeeq
where $k = \alpha_q + \alpha_g \bar{v}$.

\eAPP

\end{document}